\documentclass[cameraready]{Interspeech}

\title{SpectCount: Spectrotemporal Counting via Synthetic Signals Improves Large Audio Language Models}

\author[affiliation={1}]{Seonuk}{Kim}
\author[affiliation={1}]{Yonghyeon}{Jun}
\author[affiliation={1}]{Ju Yeon}{Kang}
\author[affiliation={1}]{Jimin}{Hong}
\author[affiliation={1}]{Yoonhyeong}{Lee}
\author[affiliation={1}]{Nam Soo}{Kim}

\address{
    $^1$ Department of Electrical and Computer Engineering and INMC, \\
    Seoul National University, Seoul, South Korea
}

\email{\{sukim, yhjeon, jykang, jmhong, yhlee\}@hi.snu.ac.kr, nkim@snu.ac.kr}

\keywords{large audio language model, spectrotemporal perception, synthetic data}

\usepackage{comment}

\usepackage{mdframed}
\usepackage{tcolorbox}
\tcbuselibrary{listingsutf8, skins}
\usepackage{tabularx}
\newcolumntype{C}{>{\centering\arraybackslash}X}

\begin{document}

\maketitle

\begin{abstract}
Large audio language models (LALMs) extend large language models with an audio encoder and large-scale audio data. However, the scarcity of high-quality annotated audio data remains a fundamental bottleneck for scaling. Through probing signal detectability analysis, we identify fine-grained spectrotemporal perceptual weaknesses in a foundation LALM. To address these challenges, we propose Spectrotemporal Counting (SpectCount), a data-efficient fine-tuning approach based on fully synthetic audio signals generated on-the-fly, without relying on real-world audio, annotations, or pretrained generative models. SpectCount not only resolves the observed weaknesses but also improves performance on diverse auditory benchmarks spanning sound, music, and speech, unseen during fine-tuning. These results suggest that weakness-targeted synthetic signals provide a data-efficient path toward enhanced auditory understanding capabilities in LALMs.
\end{abstract}

\section{Introduction}

Recent advances in large language models (LLMs) have enabled multimodal perception, extending their capabilities beyond text to audio, visual, and other modalities~\cite{vaswani2023attentionneed, wu2024nextgpt}. In the auditory domain, large spoken language models (LSLMs) integrate speech encoders with LLM backbones to support speech-centric tasks~\cite{cui-etal-2025-recent, 11278041, shon24_interspeech, aggarwal25_interspeech, kang25_interspeech}, and large audio language models (LALMs) build upon this approach to cover a broader spectrum of acoustic modalities, including environmental sounds and music, enabling more general auditory understanding~\cite{yang-etal-2025-towards-holistic, ghosh2025audio, chu2024qwen2audiotechnicalreport}.

Despite this progress, recent auditory benchmarks reveal that even foundation LALMs trained on large-scale annotated audio data still lag behind human-level performance~\cite{sakshi2025mmau, wang2026mmsu}. To overcome this limitation, researchers have explored chain-of-thought audio reasoning~\cite{yang25g_interspeech, diao-etal-2025-soundmind, zhifei-etal-2025-audio, wu2026echo}, more informative supervision signals~\cite{kuan25_interspeech, DBLP:journals/corr/abs-2511-11039}, and inference-time strategies~\cite{rong2025audiogeniereasonertrainingfreemultiagentframework, lee2025audio, taheri2025sarlmsymbolicaudioreasoning}. However, these approaches require large amounts of annotated real-world audio data, which are costly to obtain and subject to privacy and licensing constraints.

\begin{figure}[t]
  \centering
  \includegraphics[width=\linewidth]{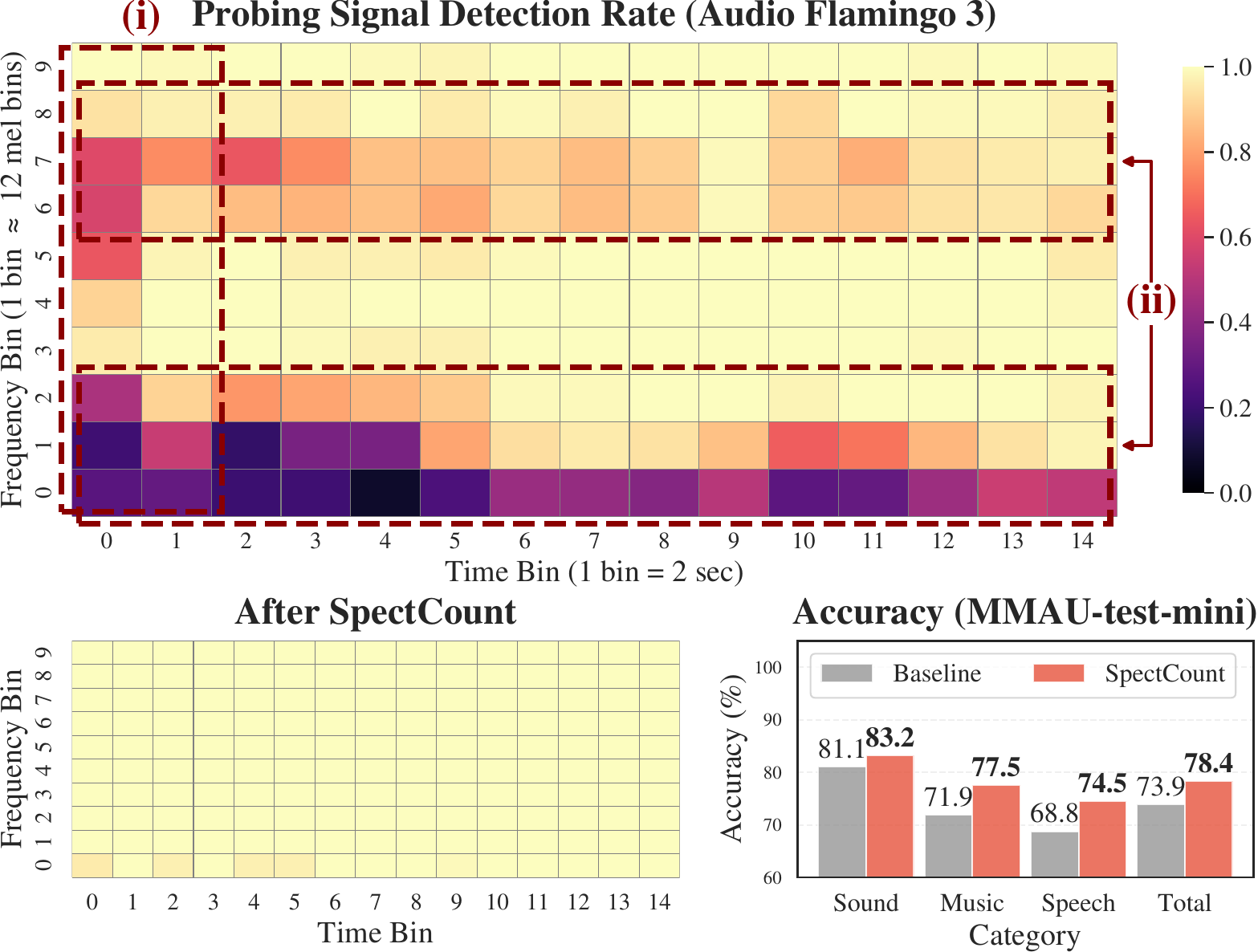}
  \caption{Probing signal detectability analysis and effects of SpectCount. The upper panel reveals two distinct weaknesses of the baseline LALM: (i) failure to recall signals appearing early in the audio, and (ii) insensitivity to specific frequency ranges. The lower panel shows the effects of SpectCount: (left) improved detection rates across the spectrotemporal space, and (right) generalization to broader auditory understanding tasks.
 }
  \label{fig:figure1}
\end{figure}

To address these challenges, one promising direction is the use of synthetic audio as an alternative data source. However, existing approaches typically use synthetic data only to supplement real-world data for specific tasks~\cite{mizumoto25_interspeech, ghosh2025synthio}, or rely on generative models that themselves require large amounts of real-world data for pretraining~\cite{ronchini2024synthetic, feng24b_interspeech}. These limitations highlight the need for more data-efficient approaches that generalize across diverse auditory tasks for LALMs~\cite{minixhofer25_interspeech, kuan2025alignment, 10890881}.

\begin{figure*}[t]
  \centering
  \includegraphics[width=\linewidth]{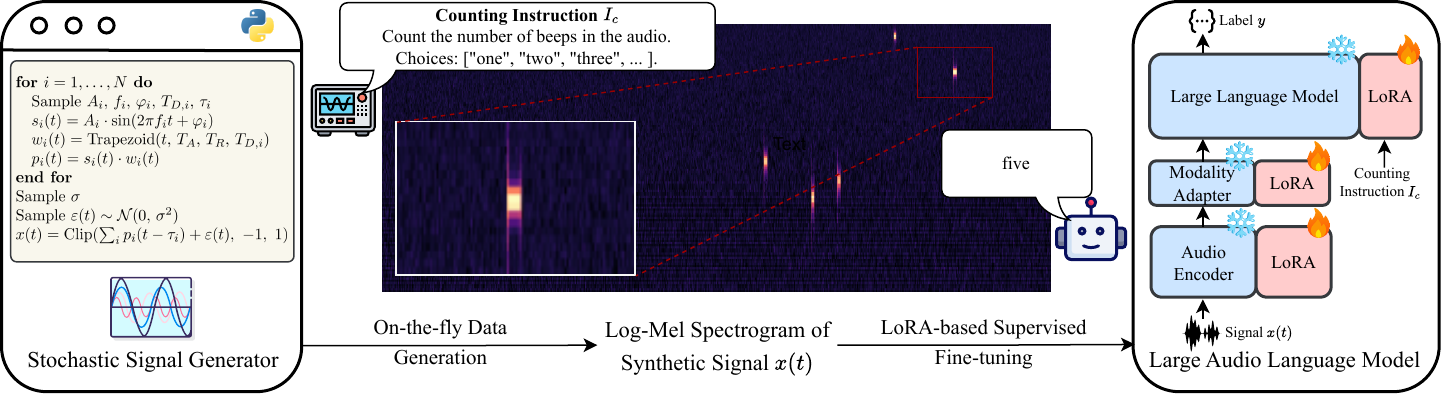}
  \caption{Overview of SpectCount.}
  \label{fig:figure2}
\end{figure*}

In this paper, we propose \textbf{Spectrotemporal Counting (SpectCount)}, a data-efficient fine-tuning approach to enhance the performance of LALMs through fully synthetic data designed to precisely target spectrotemporal perceptual weaknesses of LALMs. The upper panel of Figure~\ref{fig:figure1} provides the motivation behind our method. We probe Audio Flamingo 3~\cite{ghosh2025audio}, a state-of-the-art open-source LALM, by testing its ability to detect millisecond-scale probing signals randomly placed across the spectrogram using an instruction: \textit{Is there any short sound in this audio? Answer yes or no.} The results reveal that even a strong foundation model struggles to perceive fine-grained details within certain regions of the spectrotemporal space.

Motivated by this observation, we design synthetic signals aimed at addressing these spectrotemporal weaknesses. Specifically, the synthetic signals consist of short pulses at diverse frequency and temporal positions, each representing a fine-grained acoustic event. When visualized as a spectrogram, these pulses appear as discrete dot-like patterns along the time and frequency axes, and we train the model to count such pulses, as illustrated in Figure~\ref{fig:figure2}. Through this counting objective, the model learns to detect and aggregate fine-grained spectrotemporal information. Notably, the synthetic signals are generated on-the-fly using algorithmic rules, eliminating the need for real-world recordings, annotations, or pretrained generative models.

SpectCount fine-tunes LALMs on this task, largely resolving the previously observed spectrotemporal weaknesses, as shown in the lower-left panel of Figure~\ref{fig:figure1}.
We find that these improvements generalize to diverse auditory benchmarks spanning sound, music, and speech modalities unseen during fine-tuning, including MMAU~\cite{sakshi2025mmau}, MMAR~\cite{ma2025mmar}, MMSU~\cite{wang2026mmsu}, and AIR-Bench~\cite{yang-etal-2024-air}, as shown in the lower-right panel of Figure~\ref{fig:figure1}. These results demonstrate that the audio understanding capabilities of foundation LALMs can be meaningfully enhanced exclusively through synthetic signals, without any real-world data. We summarize our contributions as follows:
\begin{itemize}
\item We identify fine-grained spectrotemporal perceptual weaknesses in a foundation LALM through probing signal detectability analysis.
\item We propose SpectCount, a data-efficient fine-tuning method that directly targets these weaknesses using fully synthetic signals generated on-the-fly, requiring no real audio, annotations, or generative models.
\item We demonstrate that SpectCount resolves the identified spectrotemporal weaknesses and generalizes to improve performance on broad auditory understanding benchmarks across unseen domains.
\end{itemize}

\section{SpectCount}

SpectCount synthesizes training data $\mathcal{D} = \{(x_j(t), y_j)\}_{j=1}^{M}$, generated on-the-fly, where the model learns to count pulses representing fine-grained acoustic events scattered across the time–frequency space, requiring detailed spectrotemporal detection and aggregation abilities. Each signal $x_j(t)$ consists of $N$ superposed pulses ($N \sim \mathcal{U}\{1, N_{\max}\}$), mapped to a textual count label $y_j$. LALMs are fine-tuned on this data via Low-Rank Adaptation (LoRA)~\cite{hu2022lora} with a counting instruction $I_c$ that prompts the model to count the pulses within each signal. An overview of SpectCount is provided in Figure~\ref{fig:figure2}.

\subsection{Stochastic signal generation of SpectCount}

Each elementary pulse $p_i(t)$ is modeled as a sinusoid:
\begin{equation}
p_i(t) = A_i \cdot \sin(2\pi f_i t + \phi_i) \cdot w_i(t),
\end{equation}
where $\phi_i \sim \mathcal{U}(0, 2\pi)$ is the initial phase.

The trapezoidal window $w_i(t)$ is defined as:
\begin{equation}
w_i(t) = \begin{cases} 
t / T_A & 0 \le t < T_A \\
1 & T_A \le t < T_{D,i} - T_R \\
(T_{D,i} - t) / T_R & T_{D,i} - T_R \le t < T_{D,i} \\
0 & \text{otherwise}
\end{cases},
\end{equation}
where $T_{D,i}$ denotes the duration of the $i$-th pulse, with $T_A$ and $T_R$ representing the attack and release durations, respectively. This windowing mitigates spectral leakage from temporal discontinuities, ensuring signal energy remains concentrated within the target frequency bands.

Frequency $f_i$ is sampled uniformly from the center frequencies $\mathcal{F}$ of a $C_{\text{mel}}$-channel Mel-filterbank, duration follows $T_{D,i} \sim \mathcal{U}(T_{\min}, T_{\max})$, and amplitude follows $\log A_i \sim \mathcal{U}(\log \alpha_{\min}, \log \alpha_{\max})$. This stochasticity in signal generation promotes diversity in the training data.

Each signal $x(t)$ is synthesized as the superposition of $N$ pulses and additive white Gaussian noise $\epsilon(t) \sim \mathcal{N}(0, \sigma^2)$:
\begin{equation}
x(t) = \text{Clip} \left( \sum_{i=1}^{N} p_i(t - \tau_i) + \epsilon(t) \right),
\end{equation}
where $\tau_i$ is the pulse time offset, $\sigma$ is the noise level sampled as $\log \sigma \sim \mathcal{U}(\log \beta_{\min}, \log \beta_{\max})$, and the waveform is clipped to $[-1, 1]$ to prevent numerical overflow.

Each pulse time offset $\tau_i$ is sampled from $\mathcal{U}(0, T_{\text{total}})$ and accepted only if it maintains a $T_{\text{gap}}$ margin from all previously placed pulses and ends before $T_{\text{total}}$.

\begin{table*}[t]
    \centering
    \caption{Accuracy (\%) on auditory understanding benchmarks. Reported baseline scores are cited from their original papers.}
    \label{tab:main_results}
    \footnotesize
    \setlength{\tabcolsep}{0pt}
    \begin{tabular*}{\textwidth}{@{\extracolsep{\fill}} lcc cccc cccc ccc} 
        \toprule
        \textbf{Model} & \textbf{Size} & \textbf{Setting} & \multicolumn{4}{c}{\textbf{MMAU-test-mini}} & \multicolumn{4}{c}{\textbf{MMAU-test}} & \multicolumn{3}{c}{\textbf{Other Benchmarks}} \\
        \cmidrule{4-7} \cmidrule{8-11} \cmidrule{12-14}
        & (trained) & & Sound & Music & Speech & \textbf{Total} & Sound & Music & Speech & \textbf{Total} & \textbf{MMAR} & \textbf{MMSU} & \textbf{AIR-B.} \\
        \midrule
        Audio Flamingo 3~\cite{ghosh2025audio} & 8.3B & \textcolor{gray}{Base (reported)} & \textcolor{gray}{79.58} & \textcolor{gray}{73.95} & \textcolor{gray}{66.37} & \textcolor{gray}{73.30} & \textcolor{gray}{75.83} & \textcolor{gray}{\textbf{74.47}} & \textcolor{gray}{66.97} & \textcolor{gray}{72.42} & \textcolor{gray}{\textbf{58.50}} & \textcolor{gray}{61.40} & \textcolor{gray}{--} \\
        & & Base (reproduced) & 81.08 & 71.86 & 68.77 & 73.90 & 77.50 & 71.53 & 68.03 & 72.36 & 52.90 & 61.92 & 64.16 \\
        & (26.2M) & \textbf{SpectCount} & \textbf{83.18} & \textbf{77.54} & \textbf{74.47} & \textbf{78.40} & \textbf{78.20} & 73.67 & \textbf{69.50} & \textbf{73.79} & 56.30& \textbf{63.18} & \textbf{64.85} \\
        \midrule
        Qwen2-Audio-Instruct~\cite{chu2024qwen2audiotechnicalreport} & 8.4B & \textcolor{gray}{Base (reported)} & \textcolor{gray}{67.27$^{\dagger}$} & \textcolor{gray}{56.29$^{\dagger}$} & \textcolor{gray}{55.26$^{\dagger}$} & \textcolor{gray}{59.60$^{\dagger}$} & \textcolor{gray}{61.17$^{\dagger}$} & \textcolor{gray}{55.67$^{\dagger}$} & \textcolor{gray}{55.37$^{\dagger}$} & \textcolor{gray}{57.40$^{\dagger}$} & \textcolor{gray}{30.00} & \textcolor{gray}{53.27} & \textcolor{gray}{--} \\
        & & Base (reproduced) & 66.67 & 57.19 & 50.75 & 58.20 & 63.37 & 52.80 & 53.17 & 56.44 & 40.10 & 48.44 & 60.17 \\
        & (25.9M) & \textbf{SpectCount} & \textbf{70.57} & \textbf{58.38} & \textbf{61.86} & \textbf{63.60} & \textbf{69.13} & \textbf{56.13} & \textbf{58.60} & \textbf{61.29} & \textbf{45.70} & \textbf{54.24} & \textbf{62.78} \\
        \bottomrule
        \addlinespace[2pt]
        \multicolumn{14}{l}{\scriptsize $^{\dagger}$ Latest reported scores from MMAU Leaderboard (v05.15.25): \url{https://sakshi113.github.io/mmau_homepage/}} \\
    \end{tabular*}
\end{table*}

\subsection{LoRA-based supervised fine-tuning for LALMs}

To enable parameter-efficient fine-tuning while preserving the knowledge of the pretrained model, we employ LoRA, where the weight update is decomposed into low-rank matrices:
\begin{equation}
W = W_0 + BA,
\end{equation}
where $W_0$ remains frozen, and the trainable matrices $A \in \mathbb{R}^{r \times k}$ and $B \in \mathbb{R}^{d \times r}$ are constrained by rank $r \ll \min(d, k)$.

The final textual response $y$ is generated autoregressively by the LLM backbone through a concatenated sequence of projected auditory tokens $z_a$ and the counting instruction $I_c$:
\begin{equation}
y = \text{LLM} \big( [ z_a ; I_c ] \big), \quad z_a = \Phi(\mathcal{E}(x(t))),
\end{equation}
where $\mathcal{E}$ denotes the audio encoder and $\Phi$ is the modality adapter that maps audio features into the LLM's latent space.

The model is optimized via cross-entropy loss calculated on the target sequence $y$:
\begin{equation}
\mathcal{L}_{CE} = - \sum_{t=1}^{|y|} \log P(y_t \mid y_{<t}, z_a, I_c).
\end{equation}

\section{Experiments}

\subsection{Implementation details}

We applied SpectCount to Audio Flamingo 3 \cite{ghosh2025audio} and Qwen2-Audio-Instruct \cite{chu2024qwen2audiotechnicalreport} using the configuration in Table~\ref{tab:params}. LoRA ($r=8$, $\alpha=16$, dropout $0.05$) was applied to all linear layers. Training was conducted on three NVIDIA RTX 4090 GPUs with a batch size of 8, using AdamW at a constant learning rate of $2 \times 10^{-4}$. Training continued until counting accuracy converged, evaluated on a held-out set of 100 samples generated using the same procedure as the training data.

\begin{table}[th]
  \caption{Configuration for signal generation.}
  \label{tab:params}
  \centering
  \footnotesize
  \setlength{\tabcolsep}{10pt} 
  \begin{tabular}{lcc}
    \toprule
    \textbf{Parameter} & \textbf{Symbol} & \textbf{Value} \\
    \midrule
    Sampling frequency           & --              & 16,000 Hz \\
    Maximum pulse count          & $N_{\max}$      & 10 \\
    Mel-filterbank channels      & $C_{\text{mel}}$ & 128 \\
    Pulse duration range         & $T_{\min}, T_{\max}$ & 40, 160$^{\dagger}$ ms \\
    Attack/Release duration      & $T_{A}, T_{R}$       & 3, 10 ms \\
    Minimum pulse interval       & $T_{\text{gap}}$     & 40 ms \\
    Total signal duration        & $T_{\text{total}}$              & 30 s \\
    Amplitude range              & $\alpha_{\min}, \alpha_{\max}$ & 0.1, 0.9 \\
    Noise level range            & $\beta_{\min}, \beta_{\max}$   & $10^{-4}, 10^{-3}$ \\
    \bottomrule
    \addlinespace[2pt]
    \multicolumn{3}{l}{\scriptsize $^{\dagger}$ For the Qwen2-Audio-Instruct model, $T_{\max} = 200$ ms was used.} \\
  \end{tabular}
\end{table}

\subsection{Evaluation benchmarks and instructions}

We saved checkpoints every 20 steps and selected the final model as the best-performing checkpoint on MMAU-test (9k). As official instructions for reproduction  were unavailable, we reproduced the reported MMAU scores of each model as closely as possible using the following evaluation instructions:

\begin{itemize}[leftmargin=0pt, label={}]
    \item 
    \begin{tcolorbox}[enhanced, colback=gray!10, colframe=gray!50, boxrule=0.5pt, fontupper=\scriptsize\ttfamily\raggedright, left=4pt, right=4pt, top=2pt, bottom=2pt,
    title={\scriptsize\sffamily\color{black} Audio Flamingo 3}, attach boxed title to top left={yshift=-2pt, xshift=6pt}, boxed title style={colback=gray!50, colframe=gray!50, boxrule=0pt}]
    \{Question\} Please choose the answer from the following options: \{Choices\}.
    \end{tcolorbox}
    \item 
    \begin{tcolorbox}[enhanced, colback=gray!10, colframe=gray!50, boxrule=0.5pt, fontupper=\scriptsize\ttfamily\raggedright, left=4pt, right=4pt, top=2pt, bottom=2pt,
    title={\scriptsize\sffamily\color{black} Qwen2-Audio-Instruct}, attach boxed title to top left={yshift=-2pt, xshift=6pt}, boxed title style={colback=gray!50, colframe=gray!50, boxrule=0pt}]
    \{Question\}\textbackslash n Choose exactly ONE answer from the options below.\textbackslash n Output MUST be one of the options EXACTLY as written (verbatim).\textbackslash n Options: \{Choices\}\textbackslash n Answer:
    \end{tcolorbox}
\end{itemize}

To evaluate generalizability, we use the following benchmarks spanning sound, music, and speech:
\begin{itemize}
\item MMAU~\cite{sakshi2025mmau}: 10k audio understanding QAs (27 tasks).
\item MMAR~\cite{ma2025mmar}: 1k audio reasoning QAs (16 tasks).
\item MMSU~\cite{wang2026mmsu}: 5k spoken language QAs (47 tasks).
\item AIR-Bench (foundation)~\cite{yang-etal-2024-air}: $\sim$19k audio QAs (19 tasks).
\end{itemize}

\subsection{Main results}

The lower-left panel of Figure~\ref{fig:figure1} demonstrates that fine-tuning with SpectCount effectively enhances the model's sensitivity to  millisecond-scale probing signals in spectrotemporal space. More importantly, Table~\ref{tab:main_results} shows that this enhancement extends to diverse auditory understanding benchmarks, achieving 6.09\% relative improvement on MMAU-test-mini and 1.98\% on MMAU-test over the Audio Flamingo 3 base model. Notably, these gains are achieved by fine-tuning solely on synthetic data using a simple counting objective, without any exposure to real-world data from benchmark-related domains.

To further validate generalizability, we evaluate the fine-tuned model on three additional auditory benchmarks. Across all three, SpectCount achieves consistent relative improvements of 6.43\% on MMAR, 2.03\% on MMSU, and 1.08\% on AIR-Bench over the base model, further supporting its generalizability to auditory understanding across diverse domains.

Additionally, extending our experiments to Qwen2-Audio-Instruct yields consistent and even larger gains across all benchmarks, achieving 9.28\% and 8.59\% relative improvements on MMAU-test-mini and MMAU-test, respectively. This demonstrates that SpectCount is not limited to a specific LALM.

\subsection{Detailed analysis}

In this section, all analyses are performed on the MMAU-test-mini (1k) with Audio Flamingo 3.

\begin{table}[t]
    \centering
    \caption{Ablation on task formulation and fine-tuned modules.}
    \label{tab:table3}
    \footnotesize
    \setlength{\tabcolsep}{10pt}
    \begin{tabular}{lc}
        \toprule
        \textbf{Setting} & \textbf{Accuracy (\%)} \\
        \midrule
        \multicolumn{2}{l}{\textit{Task formulation}} \\
        \quad Freq-axis discrimination only & 74.7 \\
        \quad Time-axis aggregation only & 77.2 \\
        \midrule
        \multicolumn{2}{l}{\textit{Fine-tuned modules}} \\
        \quad Audio encoder only & 75.2 \\
        \quad LLM backbone only & 77.0 \\
        \midrule
        \textbf{SpectCount} & \textbf{78.4} \\
        \bottomrule
    \end{tabular}
\end{table}

\subsubsection{Ablation studies}

Table~\ref{tab:table3} presents an ablation study on two core elements of SpectCount: (i) time-axis aggregation and (ii) frequency-axis discrimination. The former is ablated by a binary single-pulse detection task similar to the probing signals described earlier, which requires only detecting the presence of a pulse rather than counting multiple pulses along the time axis, and the latter is ablated by eliminating frequency diversity, training on a single frequency band. Removing either component degrades performance, with the larger drop from time-axis aggregation indicating that temporal aggregation is the more critical contributor.

We further ablate LoRA adapter placement in Table~\ref{tab:table3} by applying it exclusively to either the audio encoder or the LLM backbone. Fine-tuning only a single module degrades performance, confirming that fine-tuning all modules is beneficial. This suggests that SpectCount adapts both the audio encoder and the LLM backbone, affecting low-level acoustic representation and high-level auditory reasoning, respectively.

\begin{figure}[t]
  \centering
  \includegraphics[width=\linewidth]{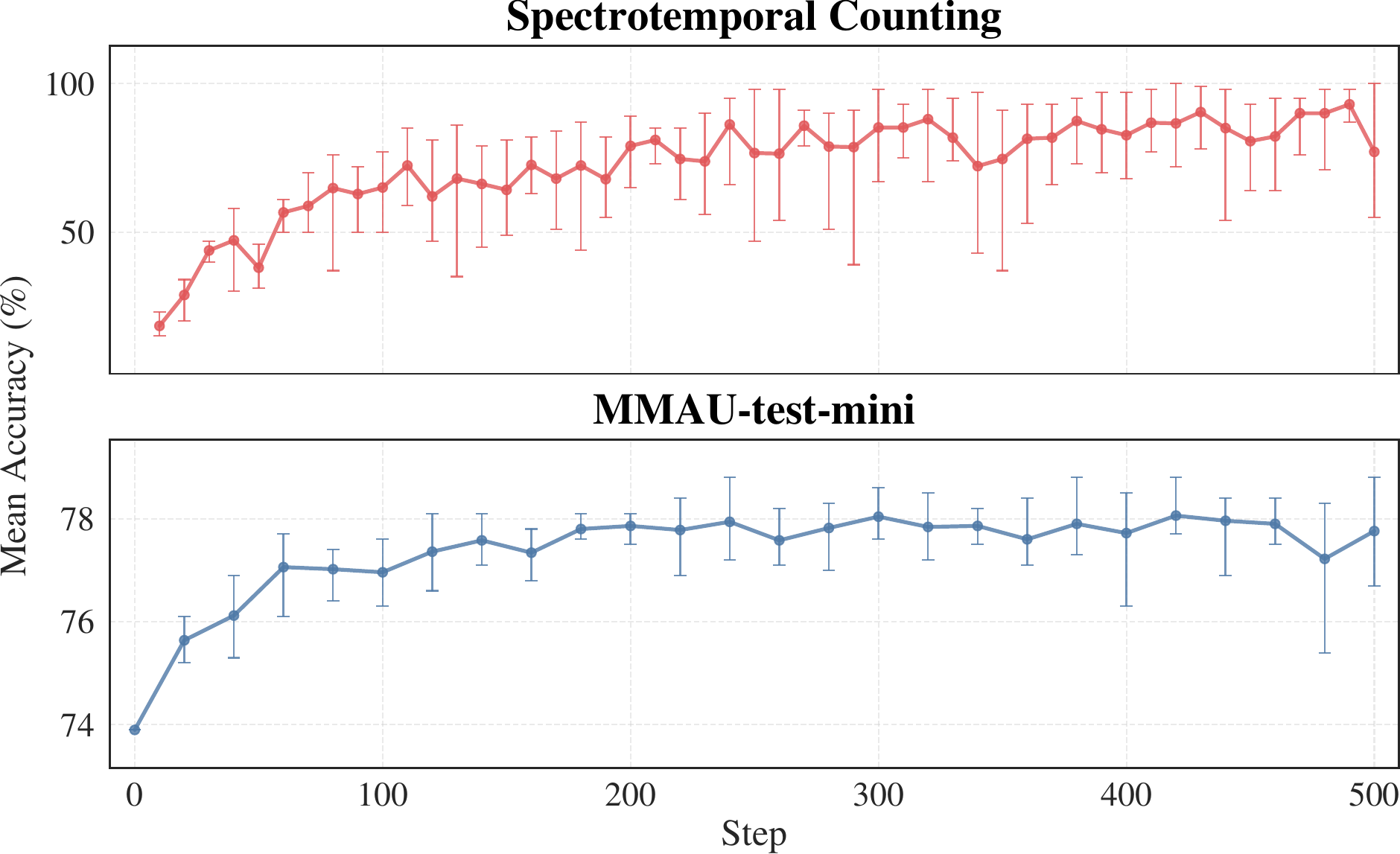}
  \caption{Accuracy (\%) curves over training steps. Error bars represent the min-max range over 5 runs.}
  \label{fig:figure3}
\end{figure}

\begin{figure}[t]
  \centering
  \includegraphics[width=\linewidth]{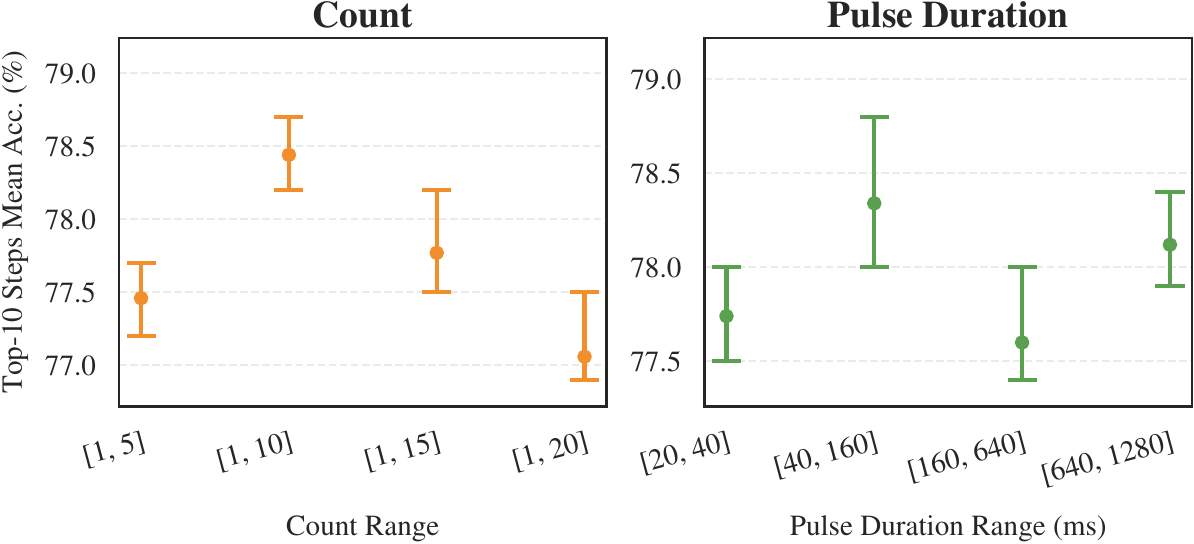}
  \caption{Impact of count and pulse duration range. Error bars represent the min-max range over 2 runs.}
  \label{fig:figure4}
\end{figure}

\subsubsection{Training dynamics}

Figure~\ref{fig:figure3} shows how spectrotemporal counting accuracy and auditory understanding accuracy progress throughout training. As training proceeds, we observe that auditory understanding accuracy improves concurrently with the acquisition of spectrotemporal counting abilities. This trend suggests that the model does not merely acquire counting as an additional isolated capability, but rather undergoes parameter adjustment that broadly benefits general auditory understanding.

\subsubsection{Impact of task difficulty}

To increase diversity of synthetic signals and prevent overfitting, signal parameters are stochastically sampled during training. Among these, count range and pulse duration are particularly important, as they directly govern task difficulty from the model's perspective, determining the number of pulses to be aggregated and the salience of each acoustic event. Specifically, a wider count range increases the memory demands for aggregation, while a shorter pulse duration produces more ambiguous acoustic events for detection. As shown in Figure~\ref{fig:figure4}, signals that are overly simple or overly complex do not yield optimal performance, suggesting that matching task difficulty to the model's learning capacity is essential. For instance, slightly longer pulse durations proved effective for Qwen2-Audio-Instruct, where reduced task difficulty led to better performance.

\subsubsection{Task-wise performance breakdown}

\begin{figure}[t]
  \centering
  \includegraphics[width=\linewidth]{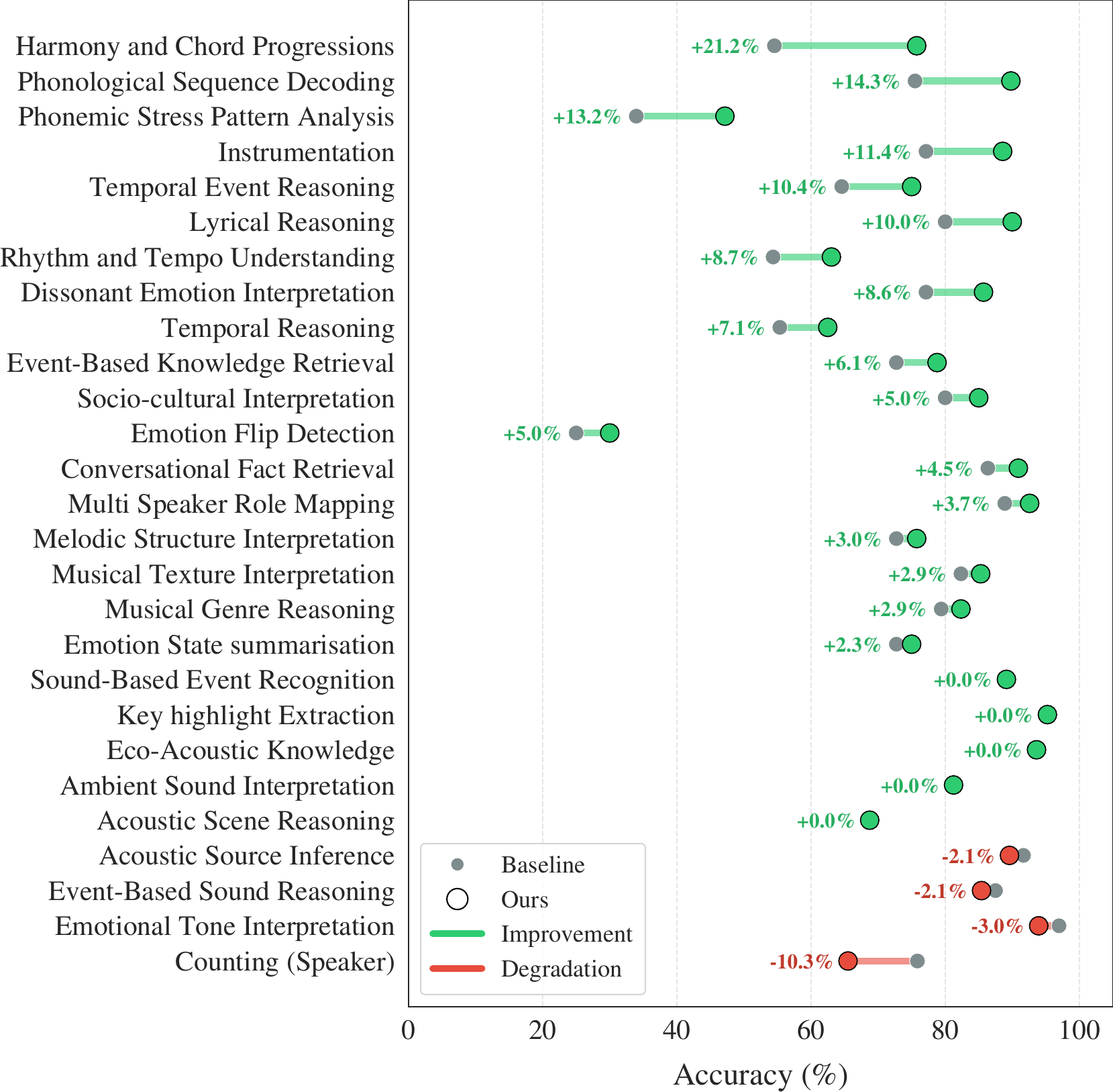}
  \caption{Accuracy (\%) improvement across auditory tasks.}
  \label{fig:figure5}
\end{figure}

Figure~\ref{fig:figure5} presents a task-level analysis of auditory understanding capabilities that benefit from SpectCount. Significant gains are observed in Harmony and Chord Progressions as well as Rhythm and Tempo Understanding, both of which require fine-grained perception to discriminate and aggregate short musical notes, precisely what SpectCount targets. Phonological Sequence Decoding and Phonemic Stress Pattern Analysis also show substantial improvements. Notably, these are achieved without any speech-related information provided during fine-tuning, suggesting that the model is capable of transferring enhanced acoustic perception to speech understanding. Furthermore, gains in Instrumentation and Temporal Event Reasoning highlight the model's enhanced ability to precisely identify the type, timing, and dominance of sound events. Conversely, the decrease in Speaker Counting performance suggests a trade-off where enhanced fine-grained perception may interfere with the recognition of global entities such as individual speakers.

\section{Conclusion}

In this paper, we propose SpectCount, a data-efficient fine-tuning method that enhances auditory perception and understanding of LALMs using fully synthetic signals. We identify fine-grained spectrotemporal perceptual weaknesses in a foundation LALM through probing analysis, and design a counting task to address these weaknesses. Experiments demonstrate that SpectCount not only resolves the observed weaknesses but also improves auditory understanding across benchmarks spanning sound, music, and speech domains unseen during fine-tuning.

\section{Generative AI Use Disclosure}
Generative AI tools were used solely for editing and polishing the English writing of this manuscript. They were not used for any core ideas or significant content.

\bibliographystyle{IEEEtran}
\bibliography{mybib}

\end{document}